%% file: main.tex
\AtBeginDocument{%
  \providecommand\BibTeX{{%
    \normalfont B\kern-0.5em{\scshape i\kern-0.25em b}\kern-0.8em\TeX}}}
\documentclass[sigconf]{acmart}

\usepackage[colorinlistoftodos]{todonotes}
\usepackage{graphicx}
\usepackage{epstopdf}
\usepackage{listings}
\usepackage{algorithm}          
\usepackage{dirtytalk} 
\usepackage{ltablex,booktabs}

\usepackage{algpseudocode}      
\usepackage{balance}
\usepackage{soul} 

\usepackage{color, colortbl}
\definecolor{Gray}{gray}{0.9}
\definecolor{LightCyan}{rgb}{0.88,1,1}
\usepackage[autostyle]{csquotes} 

\usepackage{subfigure}

\copyrightyear{2023} 
\acmYear{2023} 
\setcopyright{rightsretained} 
\acmConference[ICMI '23]{INTERNATIONAL CONFERENCE ON MULTIMODAL INTERACTION}{October 9--13, 2023}{Paris, France}
\acmBooktitle{INTERNATIONAL CONFERENCE ON MULTIMODAL INTERACTION (ICMI '23), October 9--13, 2023, Paris, France}\acmDOI{10.1145/3577190.3614158}
\acmISBN{979-8-4007-0055-2/23/10}

\begin{document}

%
\title[]{Large language models in textual analysis for gesture selection}

%

\author{Laura B. Hensel$^{1,*}$, Nutchanon Yongsatianchot$^2$, Parisa Torshizi$^2$, Elena Minucci$^1$, Stacy Marsella$^{1,2}$}
\affiliation{$^1$University of Glasgow, $^2$Northeastern University, \country{} $^*$Corresponding Author: laura.b.hensel@gmail.com}

%
\renewcommand{\shortauthors}{Hensel et al.}

%

\include{abstract.tex}

%
%


%



%

%
\settopmatter{printfolios=true}
\maketitle

\input{intro.tex}
\input{relatedwork.tex}
\input{examples.tex}

\input{approach.tex}
\input{analysis.tex}

\input{discussion.tex}

\input{conclusion.tex}

\section{Acknowledgements} \label{sec:acknowledgements}
The work in this article has been supported by EPSRC Grant Number EP/T021136/1 (LBH, SM) and UKRI CDT in Socially Intelligent Artificial Agents, Grant Number EP/S02266X/1 (EM).


\balance
\bibliographystyle{ACM-Reference-Format}
\bibliography{gesture.bib}


\end{document}

%% file: abstract.tex
\begin{abstract}
Gestures perform a variety of communicative functions that powerfully influence human face-to-face interaction. How this communicative function is achieved varies greatly between individuals and depends on the role of the speaker and the context of the interaction. Approaches to automatic gesture generation vary not only in the degree to which they rely on data-driven techniques but also the degree to which they can produce context and speaker specific gestures. However, these approaches face two major challenges: The first is obtaining sufficient training data that is appropriate for the context and the goal of the application. The second is related to designer control to realize their specific \textit{intent} for the application. Here, we approach these challenges by using large language models (LLMs) to show that these powerful models of large amounts of data can be adapted for gesture analysis and generation. Specifically, we used ChatGPT as a tool for suggesting context-specific gestures that can realize designer intent based on minimal prompts. We also find that ChatGPT can suggests novel yet appropriate gestures not present in the minimal training data. The use of LLMs is a promising avenue for gesture generation that reduce the need for laborious annotations and has the potential to flexibly and quickly adapt to different designer intents. 
\end{abstract}

%% file: intro.tex
\section{Introduction}
Human co-speech gestures have been extensively studied. Research has demonstrated that gestures convey a range of meaning and can have a powerful impact on face to face interaction \cite{kendon1997gesture,calbris2011elements,goldin2013gesture,mcneill1992hand}. Gestures positively impact the speaker's persuasiveness as well as an addressee's comprehension, recall, engagement, and trust in the speaker \cite{tversky2009embodied,tversky2012time,bavelas1994gestures}. 
However these impacts are dependent on the particular gestures being used and the context in which they occur \cite[e.g.,][]{hostetter2011}. Broadly speaking, people differ in the degree they use gestures, what types of gestures they use and the impact those gestures have \cite{kendon2004gesture,mcneill2005gesture,chu2014individual,ozer2020individuak}. This is especially clear when one considers the situation and role of the gesturer. A clinician may use gestures to establish a safe, trusted place for the interaction with a client as they discuss difficult issues \cite{foley_nonverbal_2010}. A politician may seek to be dismissive of opponents \cite{hall_hands_2016}, or arouse fear and anger \cite{casasanto2010}. A comedian may use entirely different gestures to generate laughter at their own foibles or ridicule others \cite{seizer_uses_2011}. A lack of gestures or use of inappropriate gestures can appear uncanny \cite{thepsoonthorn_exploration_2021, ostermeier_gender_1997}.

This potential power of gestures can be exploited in the creation of virtual agents. Selecting gestures for virtual agents is, however, a complex problem due to the multidimensionality of the gesture space. To solve this problem, researchers have focused on two main approaches. The first approach is to predict appropriate gestures based on analysis of the virtual human's utterances \cite[e.g.,][]{cassell2004beat, lee2006nonverbal}. Because of a lack of data sets that include both language and co-speech gestures, work in this area has sometimes exploited existing language tools~\cite{cassell2004beat,ravenet2018automating,saund2019multiple} to analyze utterance content and suggest suitable gestures. In particular, machine learning derived tools have been used to analyze prosody and parse rhetorical structure of utterances~\cite{joty2015codra} while ontology tools like Wordnet and Trains \cite{allen2020broad,miller1995wordnet} have been used to analyze lexical and semantic content. In essence, these approaches aim to generalize example gestures to semantically related utterances. However, the emphasis is increasingly shifting to use of state of the art deep learning approaches to generate gestures directly from  the utterance, in part driven by the fact that larger language+gesture corpora have become available (see \cite{nyatsanga2023} for a comprehensive review).

With any approach, the data that informs gesture selection is critical to the appropriateness of the output gestures. For example, it may be undesirable for a virtual human clinician to gesture like a politician, a comedian, or someone who rarely gestures. Similarly, it may be counterproductive to base the virtual human's gestures on an average over these different people and roles. Tailoring a virtual human's gestures to the role the virtual human inhabits and the designer's intent for the gesture's impact has potential to greatly improve the virtual human's impact \cite{salem2013err,salem2012generation}. Given this context-dependency of gestures, the central question in gesture generation becomes what gestural data can realize designer intent.

Here, we explore this question from a large language model (LLM) perspective. LLMs have the key advantage that they comprise enough data to perform textual analysis. However, the issue remains how to map that analysis to gestures, in a way consistent with design intentions. In this paper, we demonstrate that LLMs, specifically OpenAI's ChatGPT and GPT-4, are a powerful tool for selecting appropriate gestures \cite{brown2020language, openai2023technical}. We address the critical issue of design intention through the use of speaker annotations (see \cite{kipp2009gesture} for related work) consistent with  design intentions to achieve mappings from the the utterance to gesture. 

We illustrate the use of LLMs for context-relevant gesture selection based on several examples, starting with the popular but less recent model BERT (Bidirectional Encoder Representations from Transformers \cite{devlin_bert_2019}) before moving on to explore GPT in more detail. These examples demonstrate the potential power of the approach to not only select individual gestures, but also suggest novel gestures outside the data the system was provided. With this, we aim to demonstrate the transformative potential of LLM's for gesture generation research. However, we also note that the evaluation of these gestures is still a fundamental challenge in the field
\cite{wolfert2021rate,saund2021importance} and lay out a road map to formally developing the use of LLM's for gesture generation.

%% file: relatedwork.tex
\section{Related Work}
Generating co-speech gestures faces two distinct challenges: The specification problem and the animation problem \cite{neff2016}. While the specification problem deals with \textit{what} gesture should be generated, the animation problem is concerned with \textit{how} to achieve the animation of that gesture. In this paper, we focus on the specification problem. For a review of the work on the animation problem see \cite{nyatsanga2023}. The specification problem can be approached with rule-based or data-driven approaches which the following sections discuss.

\subsection{Rule-based gesture generation}
As the name suggests, rule-based approaches rely on a knowledge base along with inference procedures to determine gesture selection, where gestures are often chosen from a gesture dictionary, i.e., a corpus of gestures \cite{cassell2000}. This approach frequently relies on manual annotations of gestures and co-speech. These annotations are then used to curate a gesture lexicon as well as synthesize rules for gesture selection. For example, the Behavior Expression Animation Toolkit relies on heuristics derived from extensive behavioral research to select pre-recorded gestures based on linguistic and contextual information \cite{cassell2001}. The system uses a bespoke language tagging module and WordNet \cite{miller1995wordnet} to automatically extract this information from text. Similarly, Marsella et al. \cite{marsella2013virtual} determined the communicative function of utterances using a collection of machine learning based prosodic and syntactic analyses as well as semantic database tools. The communicative functions were then mapped to gestures based on rules codified in an indexed gesture dictionary. 

In contrast, Ravenet et al. \cite{ravenet2018automating} based their heuristics on image schemas -- recurring cognitive processes that aid in establishing meaning \cite{kendon1980gesticulation, mcneill1985so}. Example image schemas include 'big', 'full', or 'object'. The system uses WordNet to identify synonyms for in-text words that correspond to image schemas which they then map to gestures via a gesture dictionary compiled from the literature \cite{ravenet2018automating}. 

While such rule-based gesture generation systems offer a high degree of flexibility when it comes to adjusting rules for gesture-selection, and thereby changing designer intent, they have several key limitations. Firstly, their reliance on gesture dictionaries, i.e., pre-determined sets of gestures, limits the range of gestures that can be produced. As a result, the gestures lack diversity and, by extension, these systems typically only produce a limited number of different gestures for the same speech input. Secondly, generating such extensive heuristics and corresponding gesture dictionaries is time consuming, relies on expert-knowledge and is therefore not scaleable. Because of these limitations, a growing body of work is focusing instead on data-driven gesture generation.

\subsection{Data-driven gesture generation}
Early approaches to data-driven gesture generation often relied on probabilistic mapping from semantic content to gesture type \cite{neff2008gesture, ishi_speech-driven_2018}. For example, several approaches used annotated gesture data to produce gesture profiles to which they could apply statistical analyses to ultimately predict plausible gestures for new, annotated input-text \cite{kipp_gesture_2003, neff2008gesture}. Similarly, Bergmann and Kopp \cite{bergmann2009increasing} built a Bayesian network to determine whether and what type of gesture should be generated, based on variables such as the communicative action the gesture aims to achieve. While these approaches are more flexible than rule-based approaches, they still often rely on laboriously annotated data and a relatively small set of variables to apply the statistical analyses to.

Because of this, data-driven gesture generation increasingly uses machine learning techniques, and specifically deep learning \cite{nyatsanga2023, ahuja2020}. Many of these approaches use encode-decoder recurrent neural networks that learn the mapping from utterance text to gestures \cite{yoon2020, yoon2019robots, ao_rhythmic_2022}. However, these approaches often still rely on gesture lexica \cite{ao_rhythmic_2022} and in all cases, these data-driven approaches are dependent on the data they are trained on, with limited data sets available. Additionally, much of the semantic richness is often lost \cite{nyatsanga2023}. 

\subsection{Large Language Models}
In contrast, the extensive pre-training of LLMs lends itself as a natural next step in automatic gesture generation. Currently, there are multiple available LLMs, such as OpenAI’s GPTs, Meta’s LLaMA, Google’s Bard, and Anthropic’s Claude, to name a few (for a recent review, see \cite{yang2023harnessing}) \cite{brown2020language, touvron2023llama, google2023bard, anthropic2023claude, openai2023technical}. One LLM that has found much traction in the research community is BERT \cite{devlin_bert_2019}. BERT consists of transformer encoder layers and is designed for language modeling and next-sentence prediction. It has been shown to powerfully predict semantic similarities between word- and sentence-pairs \cite{mutinda_semantic_2021,peinelt_tbert_2020} and this makes it a potentially valuable tool for gesture generation as mappings from words to gestures can be generalized over synonyms \cite{fares_transformer_2022, kucherenko_gesticulator_2020}. However, BERT does not have in-built explainable AI, limiting its application to realizing designer intent. For this reason, we explore BERT only briefly before we primarily focus on models from OpenAI. 

We chose OpenAI's models for two key reasons: First, they provide easy access API and second, at the time of writing, GPT-4 is the most recent and powerful model \cite{peng2023instruction}. Specifically, we use two of OpenAI's Generative Pre-trained Transformer models or GPT: ChatGPT (\texttt{gpt-3.5-turbo}) and GPT-4 \cite{brown2020language, openai2023technical}. GPTs are an autoregressive decoder-only transformer-based large language models. They are pretrained on large data sets to predict next words and fine-tuned using Reinforcement Learning with Human Feedback to make the models align to human preferences and common tasks \cite{ouyang2022training}. One of the key capabilities of these LLMs in relation to this work is in-context learning -- their ability to learn a new task given only a few examples without having to fine-tune the model \cite{brown2020language, xie2021explanation}.

Recent work has begun to demonstrate the ability of GPTs' to approximate human cognitive processes. For example, Binz and Schulz \cite{binz2023using} studied GPT-3's decision-making, information search, and causal reasoning using common cognitive psychological tests such as the conjunction and base-rate fallacies. They found that GPT-3 can solve these problems at a similar or better than human level but fall for some common human biases. Similarly, Kosinski \cite{kosinski2023theory} examined Theory of Mind in LLMs using standard false-belief tasks and found that ChatGPT and GPT-4 can solve most of these tasks. Similarly, Bubeck et al. \cite{bubeck2023sparks} tested GPT-4 in various domains including mathematics, coding, medicine, law, and psychology, including Theory of Mind and emotion understanding. They showed that GPT-4 can perform  these tasks well and outperforms ChatGPT. Taken together, this body of works shows GPTs' promise for applications that normally require human cognitive input. 

One such application is gesture generation. GPTs' extensive textual analysis training has the potential to yield gestures that are semantically rich and flexible while not relying on manual annotations. The first foray into this approach used ChatGPT to identify the intent of an utterance, based on this their system then selected a gesture based on a pre-compiled gesture dictionary \cite{gao_gesgpt_2023}. The work demonstrated that ChatGPT can successfully aid in gesture generation but did not yet fully realize its potential which lies in ChatGPT's ability to a) suggest new gestures with little input and b) adapt gestures to designer intent.

\subsection{Gesture generation driven by designer intent}
Work that aimed to realize designer intent has often focused on producing gestures for a specific novel target, i.e., a single gesturer. For example, Ahuja et al. \cite{ahuja2022} were able to generate gestures for a novel target by adapting a pre-trained source model based on just two minutes of target data. Their model did this by learning the difference in the crossmodal grounding, i.e., binding between language and physical gestures, between the source model and the target data. In contrast, other work has focused on learning different speaker styles that can then be used in a zero-shot style transfer to generate gestures for targets not present in training data \cite{fares2023, ghorbani2023}. These approaches are often very efficient but are also tailored to generate gestures for specific targets.

%% file: examples.tex
\section{examples}
One approach to applying modern LLMs to gesture generation is to use them to replace the generalization from examples. Earlier rule-based work discussed in previous sections relied on analyses using a range of ontology based tools \cite{grady1997foundations,miller1995wordnet,allen2020broad}. Such tools can be used to analyze text to find similarities with, and thereby generalize from, a set of annotated text-gesture pairings. For example, consider the phrase "important concept". A rule based ontological approach may codify the metaphoric mapping from "important" to \textit{big} and the semantic mapping from "concept" to \textit{idea} which in turn can be metaphorically mapped to a \textit{container} gesture. As the following example shows, language models have the potential to replace this codification. 

\subsection{BERT}
The pre-trained BERT model can realize a mapping from text to gesture directly through its cosine similarity measures, even without re-training. Given some phrase-gesture example mappings, derived for example from some annotations of a speaker, and a novel utterance, BERT can measure the cosine similarity between each example utterance and a novel utterance. Additionally it provides a metric for this similarity which can be used to choose potential alternative gestures. Consider the following example where BERT gives similarity measures between \textit{important concept} and five phrases of varying relatedness.

\vspace*{0.1in}
Seed Phrase: {\em important concept}
\vspace*{-0.08in}
\begin{itemize}
\item big idea correlation: 0.861
\item great thought correlation: 0.81
\item red napkin correlation: 0.239
\item weak idea correlation: 0.469
\item silly idea correlation: 0.674
\end{itemize}

We see a strong relation between the seed {\em important concept} and {\em big idea} as well as {\em great thought}. As expected the relation to {\em red napkin} is the smallest as they are semantically unrelated. The seed's relation with {\em weak idea} as one might expect is somewhat stronger but still weaker than  {\em big idea} and {\em great thought}. In contrast, the seed's relation to {\em silly idea} is slightly stronger than {\em weak idea}, arguably because silliness has both positive and negative attributes. Critically, however, the exact reason for this lower correlation is hidden from the user at this level of probing the model.

In short, BERT's pre-trained model could substitute for the lexical, semantic, metaphoric analyses discussed above. One would expect BERT to realize a much broader generalization than an ontological approach, though this depends on the boundary condition set on the cosine similarity. A key benefit of BERT is that it provides a clear, overall metric on this mapping as the above examples show. However, the derivation of this metric is opaque and outside the designer's control. 

In contrast, the ontology based approach is more transparent as to how this mapping of a novel utterance to gesture suggestion is realized. The semantic, lexical and metaphoric elements are well-defined, which potentially supports finer grain designer manipulations. For example, specific analysis elements could be eliminated or discounted, such as mapping "important" to \textit{size}, or specific analyses could be generalized, such as generalizing abstract concepts to any concept so any object, abstract or physical could be conveyed by a container.

Whereas BERT is arguably similar to prior approaches, ChatGPT, specifically its prompting ability, provides another, novel approach to achieving generalization from example annotations while allowing for designer control through the prompting mechanism.

\begin{figure*}[h]
\centering
  \subfigure[Anything]{\includegraphics[scale=0.095]{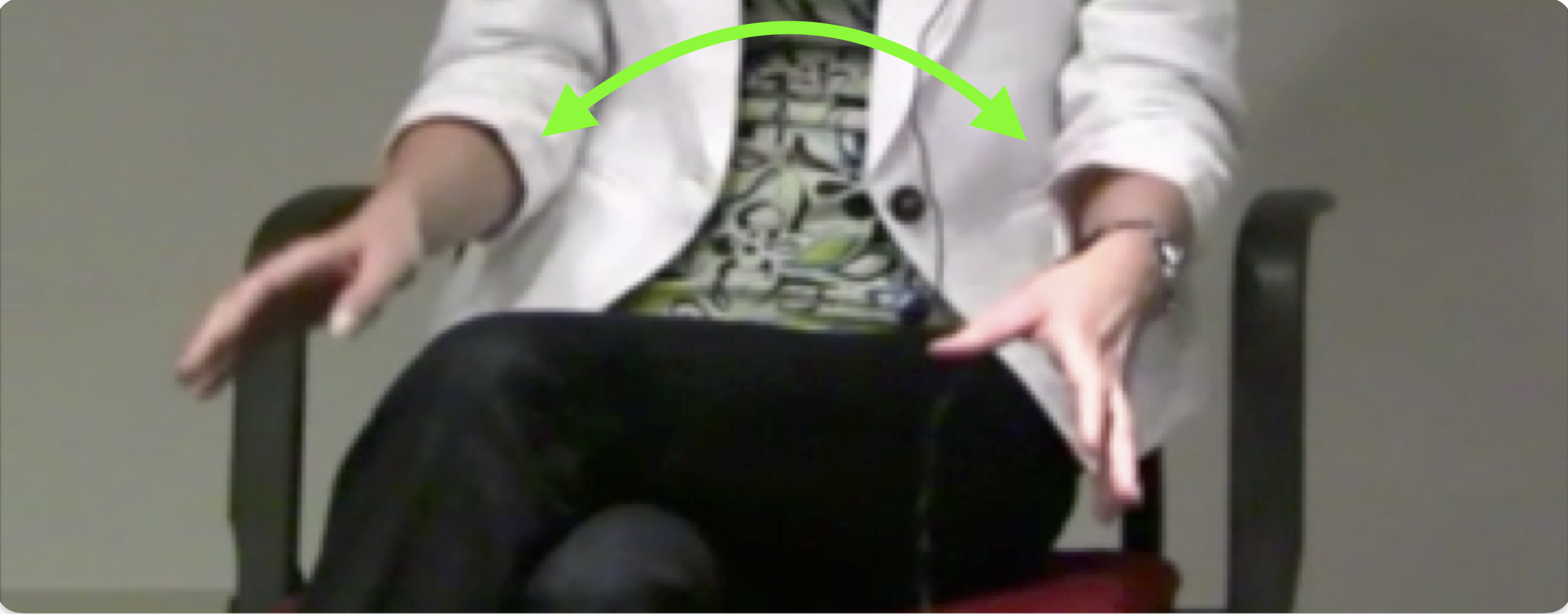}}\hfill
  \subfigure[Besides]{\includegraphics[scale=0.095]{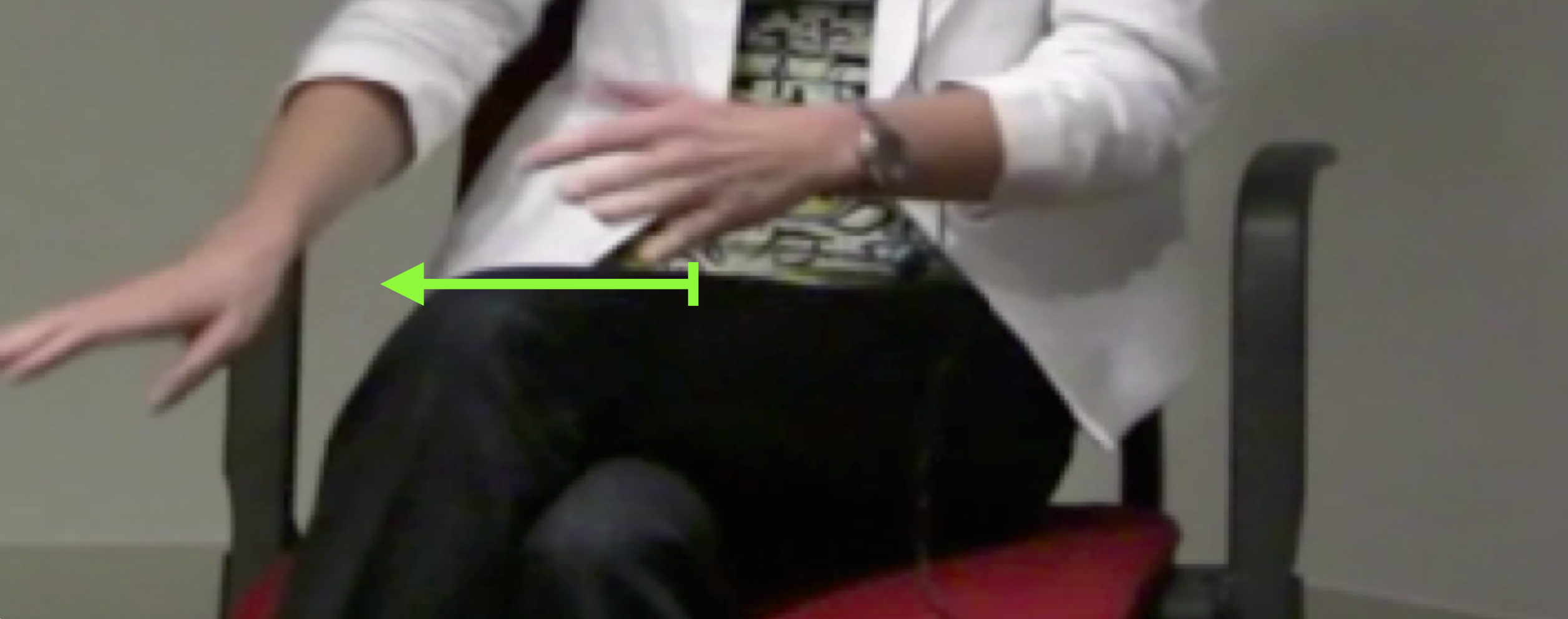}}
\caption{Ideational Unit showing a container gesture depicted a protected space followed by a sweep gesture suggesting something discarded from the container}
\label{fig:clinician_example}
\end{figure*}

\subsection{ChatGPT and GPT-4}

To illustrate the power of LLMs for gesture selection, we prompted ChatGPT and GPT-4 with text-gesture pairings which in practice would be drawn from annotations of an actual speaker or speakers. In the following example, we intentionally used opaque terms for the gesture that ChatGPT could not interpret, such as "lcg" meaning "large container gesture" and "ng" for "negation gesture". Based on just three example prompts when given the phrase "it is an excellent idea", ChatGPT responded with "lcg", shown in bold italics.

\paragraph{ChatGPT}
\begin{itemize}
    \item Example prompt: \textit{It is an important idea. lcg}
    \item Example prompt: \textit{It is a silly idea. ng}
    \item Example prompt: \textit{It is a big idea. lcg}
    \item Prompt: \textit{It is a excellent idea. }
    \item Output: {\em \textbf{lcg}}
\end{itemize}

This method is very similar to the BERT approach -- it identifies the similarity between "important", "big", and "excellent" and produces the corresponding gesture.
However, if we supply gesture descriptions that are interpretable to ChatGPT, then it produces novel gestural suggestions based on its interpretation of the gestural prompts. The following example demonstrate this. 

\paragraph{ChatGPT}
\begin{itemize}
    \item Example prompt: \textit{It is a important idea. large container gesture} 
    \item Example prompt: \textit{It is a bad idea. no gesture}
    \item Example prompt: \textit{It is a big idea. large container gesture}
    \item Prompt: \textit{It is an excellent thought.}
    \item Output: {\em \textbf{thumbs up gesture}}
\end{itemize}

ChatGPT suggests a gesture that is appropriate for illustrating the utterance but was never prompted: {\em{"thumbs-up gesture"}}. Such novel suggestions raise the question as to the degree to which the prompting is reflecting an underlying relation between utterance and gesture. To illustrate the depth of GPT's ability in this regard, consider an utterance sequence adapted from an interaction between a clinician and woman suffering from PTSD. This utterance and gestural sequence is adapted\footnote{We did not use the exact utterance on one hand to make the context explicit, i.e., a clinician talking to a client, and additionally to remove dysfluencies \cite{lhommet2014metaphoric}.} from an 90 minute corpus in which a real world clinician is going through a PTSD protocol with a woman actor playing the role of a client suffering from PTSD. She is talking to the clinician because of her husband who wants her to change behavior. A key moment in he interaction is as follows:

The clinician is asking a client whether (1) there is anything the client wants to work on, adding the proviso of (2) besides what her husband wants. She uses the following gestures: As she says says (1) "anything", she performs a container gesture suggesting a set of all the possible topics they could discuss. Specifically, this container gesture depicts a domed region of space, perhaps suggesting a protected space (see Figure \ref{fig:clinician_example}). Next,
she clarifies what topic she does not want to discuss, (2; \enquote{besides what he wants}) using a gesture that sweeps to the right, suggesting removing that topic from the container formed by the previous gesture. The "he" is a reference to the client's husband. See Figure \ref{fig:clinician_example}. These gestures constitute what Calbris \cite{calbris2011elements} calls an ideational unit, a sequence of related gestures. Such gestures tend to set up gestural space in a consistent fashion, such as depicting sets of abstract things and then removing or putting items into the set. 

To explore GPT's ability to form ideational units, we first prompted ChatGPT to provide an explanation for the use of the container gesture in the clinician example. This yielded the following explanation that is consistent with a prior analysis of the clinician's gesture done by gesture researchers \cite{devault_simsensei_nodate}:

\paragraph{ChatGPT}
\begin{itemize}
    \item Prompt: \textit{The clinician said we can talk about anything. She used a container gesture when she said anything.}
    \item ChatGPT: \textit{The clinician's use of a container gesture when saying "anything" may suggest that she wants you to feel safe and comfortable to discuss any topic that comes to mind, but also to keep those topics contained within the therapeutic space. The container gesture is a nonverbal cue that can signal a sense of containment and boundaries, creating a safe space for exploration and self-discovery.}
\end{itemize}

Thus, ChatGPT captured the relation of the container to the possible topic. The response also unexpectedly captured the notion of a protected space which aligns with the specialized dome-shaped container gesture used by the clinician which we did not include in the prompt. GPT-4 yielded similar results but did not capture the subtlety of a  "safe space" as explicitly. It instead suggested a container gesture to show inclusiveness and make the patient comfortable.

Importantly, the following example shows that GPT-4 can provide a good prediction of what gesture the clinician could perform next consistent with the gesture sequence the clinician actually used (ChatGPT yielded similar results) . This suggests an ability to construct an ideational unit.

\paragraph{GPT-4}
\begin{itemize}
    \item Prompt: \textit{The clinician said "We can talk about anything" and then she added "besides what your husband wants us to talk about." She used a container gesture to illustrate "talk about anything." To illustrate "except what your husband wants us to talk about", what gesture might she use after the container gesture.}
    \item GPT-4: \textit{After using the container gesture to illustrate "talk about anything", the clinician could use a swiping or brushing away gesture with one hand to represent "except what your husband wants us to talk about". This gesture would visually convey the idea of removing or excluding a specific topic from the otherwise open and inclusive container of conversation topics.}
\end{itemize}

Finally, we prompted GPT-4 to visualize the entire action:

\begin{itemize} 
\item Prompt: \textit{Can you visualize it in some way?}
\item GPT-4: \textit{Sure! Here's a step-by-step visualization of the gestures:}
\begin{enumerate}
\item The clinician starts by using a container gesture to illustrate "talk about anything." She holds her hands in front of her, palms facing each other, and forms an imaginary circle or container in the air.
\item To illustrate "besides what your husband wants us to talk about," she then moves one hand, palm facing down, from one side of the imaginary container to the other, as if swiping or brushing away a specific topic.
\item The clinician's facial expression could also change to emphasize the exclusion, such as raising an eyebrow or giving a slight smile to show that she's aware of the exception and is still open to discussing other topics.
\end{enumerate}
    \item This combination of gestures and facial expressions would help to visually convey the message that all topics are open for discussion, except for the specific topic the husband wants them to talk about.
\end{itemize}

Note GPT-4 has gone beyond the hand gestures to suggest facial actions. These examples illustrate that there is important, and often surprising, potential for gesture selection and analysis in LLMs. The next sections explore this potential more systematically.

%% file: approach.tex
\section{Approach}

In this section, we detail two approaches to investigating the use of LLMs to a) predict gestures from annotated examples and b) suggest gestures without example prompts. Note that these do not form an exhaustive exploration of using LLMs for gesture selection but are aimed at giving initial insight into the power of these tools. In the discussion, we set out a road map for developing these approaches and the use of LLMs in gesture generation. Note that all analyses in this paper are based on GPT versions from March 2023. 

\subsection{Data collection}

To demonstrate GPT's potential for gesture selection, two researchers jointly annotated the first four minutes of Barack Obama's 2020 speech to the Democratic National Convention. Specifically, we focused on three gesture categories that were easily identifiable and semantically meaningful: sweep, span, and container. In total, we identified thirty-seven such gestures. We further distinguished between six unique physical gesture types: span, container, and four types of sweeps with different orientations of the palm (up, down, in, and forward). We also classified these gestures in terms of the semantic meaning conveyed, such as temporal span, negative sweep, and enumeration sweep. The final dataset contained seventeen unique semantic gestures (semantic gesture description + gesture category) and fifteen unique semantic gesture descriptions. Code and data are available at:
\url{https://osf.io/c82tq/?view_only=4e0c9c1923cc432abd2c82d0ed8f05c8}

\subsection{GPT gesture prompting}

For this work, we used ChatGPT and GPT-4. In all of our experiments, we set the temperature parameter to 0, making the outputs mostly deterministic to maximize the replicability of our results. All other parameters were set to default. 

To elicit gesture suggestions, we prompted ChatGPT and GPT-4 in the following ways. First, we provided each model with a number of examples (2, 4, 6, or 33, i.e., leave-one-out) for each gesture category. For example, we provided two container gestures, two sweep gestures, and two span gestures. Each prompt was structured in three parts: 

\begin{enumerate}
  \item The following contextual statement:
    \begin{displayquote}
    Barrack Obama is giving a speech at the Democratic National Convention.
    \end{displayquote}
  \item Examples of gestures Obama used. Each example consisted of three elements: The segment of the speech the utterance occurred in, the utterance associated with the gesture, and the gesture. For example:

    \begin{displayquote}
    He said \say{I want to talk as plainly as I can about the stakes in this election because what we do these next 76 days will echo through generations to come.} When he said \say{these next 76 days}, he used the following gesture: \say{span}. 
    \end{displayquote}
  \item The target utterance excluding the gesture. The models filled in the gesture at this point.
\end{enumerate}

We investigate four gesture specifications: Gesture category (span, sweep, container), physical gesture descriptions (e.g., sweep with palm facing up/down), semantic gesture descriptions (e.g., negative sweep), and semantic description only (e.g., negative). 

The semantic specification in particular yielded a variety of gestures. We summarise an illustrative selection of these in Table \ref{tab:1} for the reader's convenience.
\begin{table*}[h!]
    \centering
    \begin{tabular}{|p{1.56in}|p{2.5in}|p{2.5in}|}
    \hline
        \textbf{Original gesture} & \textbf{ChatGPT} & \textbf{GPT-4} 
        \\ \hline
        \textbf{inclusive span} & inclusive span/sweep & inclusive span/sweep, compare and contrast
        \\ \hline
        \textbf{temporal span} & temporal span, spatial span, forward span, positive span & temporal span, negation span, counting on fingers or tapping on wrist, forward sweep, guiding motion, continuation, positive sweep, process sweep
        \\ \hline
        \textbf{positive sweep} & upward sweep, emphatic sweep, temporal span & upward sweep
        \\ \hline
         \textbf{negative sweep} & negative/negation span/sweep, inclusive span, negative head shake & negative sweep, downward sweep, negation span, negation or dismissive wave, shrug or open palm, uncertain span
        \\ \hline
    \end{tabular}
    \caption{ChatGPT and GPT-4 gesture suggestions for semantic specifications}
  \label{tab:1}
\end{table*}

Finally, to explore GPT's ability to suggest appropriate gestures without any additional user input, we input each utterance prompt without gesture examples. An example includes:
\begin{displayquote}
Barrack Obama is giving a speech at the Democratic National Convention. He said \say{I want to talk as plainly as I can about the stakes in this election because what we do these next 76 days will echo through generations to come.} When he said \say{these next 76 days}, what gesture did he use?
\end{displayquote}

%% file: analysis.tex
\section{analysis}
\begin{figure*}[htbp]
\centering
\includegraphics[width = 1\textwidth]{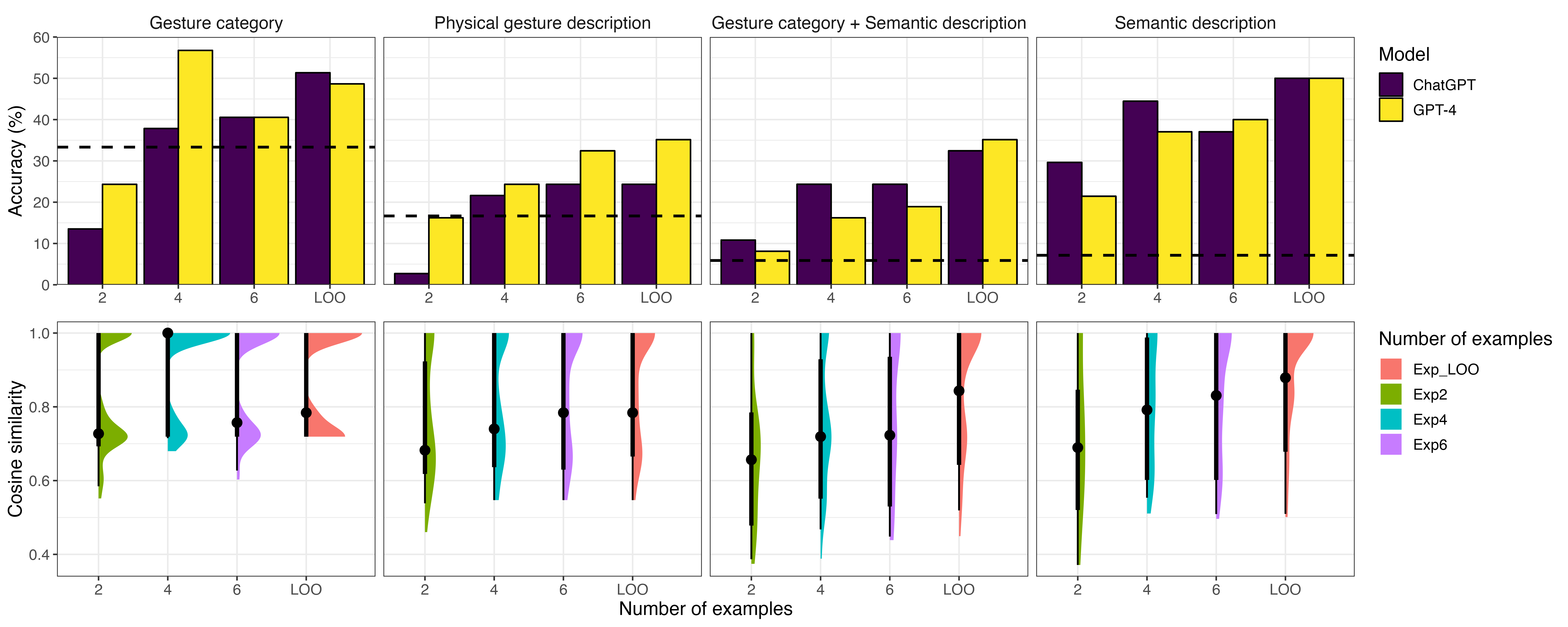}
\caption{Prediction performance for four gesture specifications. Top: Prediction accuracy (percent correct). Bottom: Cosine similarity between true and predicted gesture (GPT-4).}
\label{fig:main_results}
\end{figure*}

To explore LLMs' performance at predicting context-specific gestures, that is, gestures produced by Barack Obama during his speech, we first calculated prediction accuracies for the four gesture specifications (gesture type, physical gesture description, semantic gesture, semantic only) and for each level of example numbers (2, 4, 6, and leave-one-out). Note, that for semantic specifications, we based this analysis on the first part of the semantic descriptor (e.g., 'negat', 'inclu') to avoid bias from slight deviations in spelling. Next, we measured pairwise cosine similarity between each original gesture and GPT-4's suggestions based on word embeddings derived using BERT (ChatGPT excluded for brevity, results are similar). This analysis accounted for cases in which GPT's description of the gesture differed from the annotation, therefore appearing inaccurate, but nevertheless showed high semantic similarity. Figure \ref{fig:main_results} shows the results. 

Specifically, in the top row, each plot shows the prediction accuracy (y-axis) for each level of examples (x-axis), where 'LOO' denotes leave-one-out, for ChatGPT (purple) and GPT-4 (yellow), separately. The dotted line indicates chance performance (1/number of unique gesture types of that category in the dataset). In the bottom row, each plot shows the distribution of pairwise cosine similarities, ranging from 0 (very different) to 1 (very similar), for each level of examples (x-axis) for GPT-4.

Results show that both models predicted the correct gesture category and the correct physical gesture description with above chance accuracy when provided with four or more example prompts. Moreover, both models correctly predicted gesture types with semantic description and on only semantic descriptions with above chance accuracy from as little as two example prompts. Across all four gesture specifications, increasing the number of example prompts improved prediction performance and we observed the highest performance during gesture category and semantic description prediction. However, two examples appear to be too few to predict specific physical gestures. Instead, the models frequently generated gesture descriptions that did not map directly onto the gesture types we examined, thus causing the low performance. Notably, GPT-4 outperformed ChatGPT when predicting gesture categories and physical descriptions but not when predicting specifications involving semantic information such as 'negative' or 'inclusive'. 

\begin{figure}[htbp]
\centering
\includegraphics[width = 1\columnwidth]{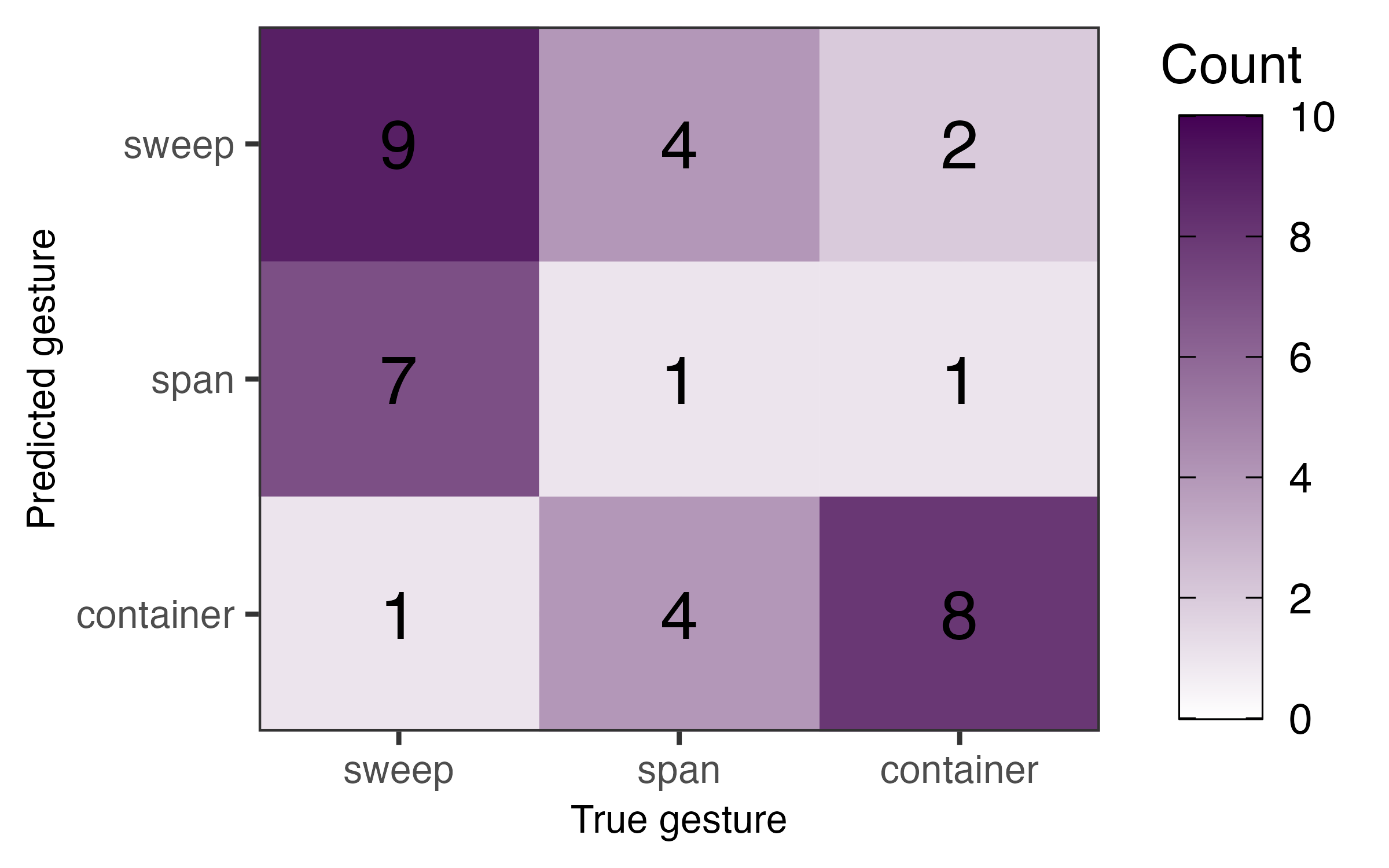}
\caption{Confusions of gesture category prediction by GPT-4.}
\label{fig:confusion_matrix}
\end{figure}

Figure \ref{fig:confusion_matrix} shows that misclassifications were primarily driven by confusions between span and sweep gestures. GPT-4 performed particularly poorly in this regard (shown in Figure \ref{fig:confusion_matrix}). This indicates that the models used these gestures interchangeably. 


\begin{figure}[htbp]
\centering
\includegraphics[width = 1\columnwidth]{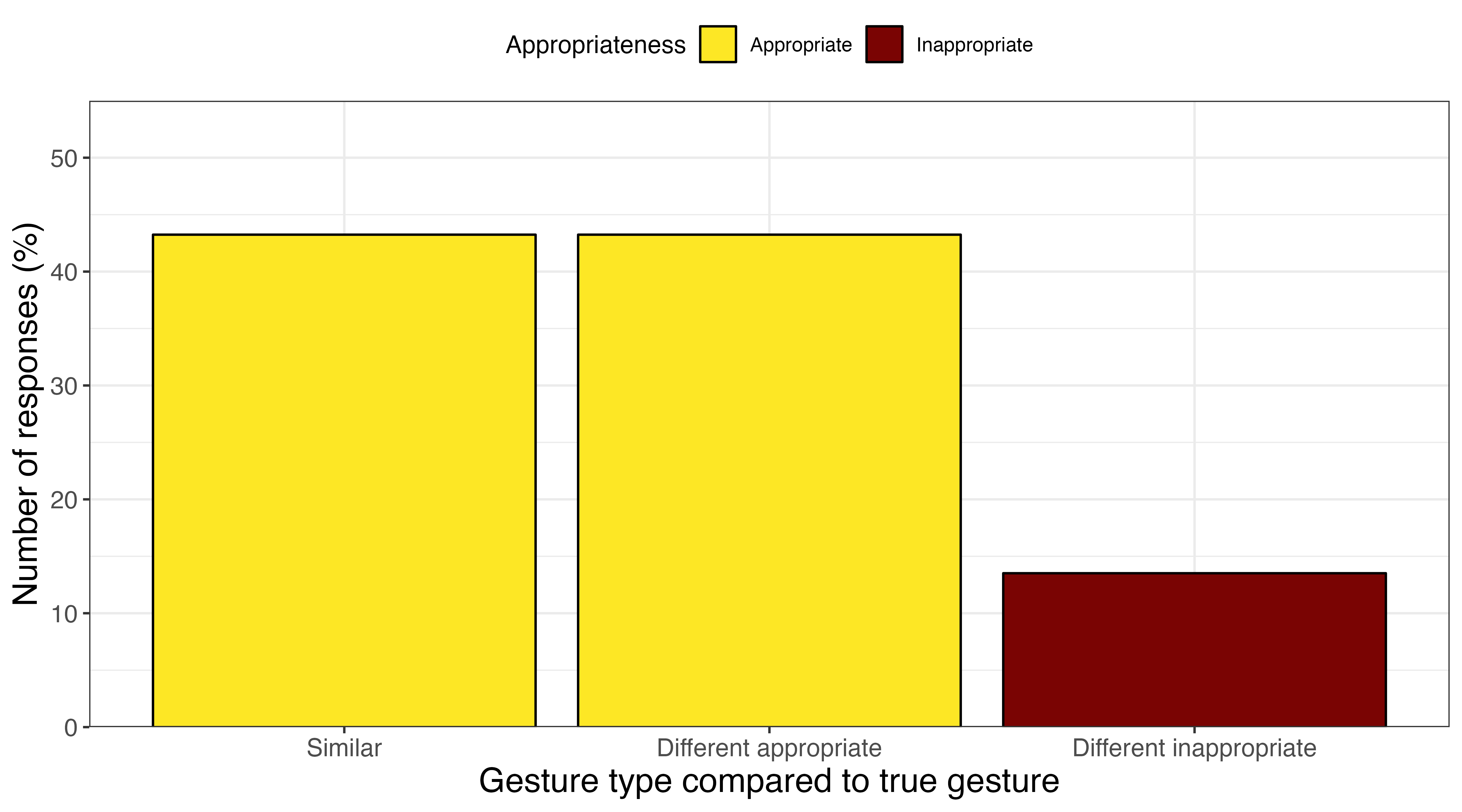}
\caption{Appropriateness of gestures suggested by GPT-4 without example prompts.}
\label{fig:appropriate}
\end{figure}

Finally, to explore GPT-4's ability to generate appropriate co-speech gestures without any example input, we categorized the models' gesture output when no examples were provided based on its appropriateness. Specifically, we applied one of the following labels to each output gesture: Similar (to the original gesture), Different appropriate (the gesture was different but still appropriate in the context of the speech), Different inappropriate (the gesture was different and not appropriate in the context of the speech), No gesture (the model did not supply a gesture). Two researchers categorised all 37 output gestures of GPT-4, for brevity, and discussed any ambiguous cases until an agreement was reached. Figure \ref{fig:appropriate} shows the resulting appropriateness levels. Red denotes inappropriate suggestions. 

Results showed that in 43.2\% of cases GPT-4 suggested a gesture  similar to the true gesture. Importantly, when the produced gesture was not similar to the true gesture, in most cases (43.2\%), the novel gesture was nevertheless appropriate in the context of Obama's speech. We found GPT-4 only produced inappropriate gestures in 13.5\% of cases. 

Notably, the model not only made appropriate suggestions for gestures but frequently supplied relevant and appropriate justifications for these suggestions. For example, GPT-4 described the gesture for 'unleashed' as "releasing something or setting free". This description very closely mirrored the actual gesture observed which was a quick motion of the right hand from the palm facing toward the body to the palm facing upward and the hand moving forward.

%% file: discussion.tex
\section{Discussion}
Here, we illustrated several key benefits of using LLMs for gesture selection. We showed that LLMs (1) can predict gestures used in a real-world context with above chance accuracy from as few as two example prompts and (2) can generalize over annotations to suggest novel, previously unseen gestures that are context-appropriate. Combined, these results demonstrate a key benefit of using LLMs for gesture selection -- designer control, which is enabled by the selection of the speaker and context. For example, if the design goal for a virtual human is a clinical application where nonverbal behavior can play a critical role \cite{HALL199521}, the annotation of a trained clinician with highly effective gesturing skills interacting in a clinical session provides an ideal seed. Indeed, the clinician example highlighted the LLM's powerful ability to adapt to specific contexts as it appeared sensitive to the clinical context when providing an interpretation of the clinician's gestures.

Note that we provided the models with very few example prompts and required outputs that matched relatively strict criteria (e.g., 'sweep', 'palm down sweep', 'negat sweep', 'negat'; depending on gesture specification). Additionally, the differentiation between spans, sweeps, and containers is nuanced because each suggests a region in space. This further adds to the difficulty of the task. Finally, the models were not fine-tuned to achieve gesture selection. Taken these challenges into account, the difficulty of the task should not be underestimated and the observed prediction accuracies of well above chance represent an impressive performance. Moreover, both models not only predicted gesture types and physical gesture descriptions but also semantic gesture specifications. This is especially impressive as we did not provide the models with information about different gesture types or how they relate semantically. Nevertheless, the models were capable of suggesting plausible gestures.

However, a key challenge to applying LLMs to gesture generation is the annotation process itself. For one, manual annotation is laborious. Additionally, designing a gesture naming scheme can be difficult due to the wide variety of real-world gestures and the lack of an agreed upon coding system for gestures, unlike the facial action coding system used in facial expression research\cite{ekman1978facial}. Any coding scheme used needs to reliably distinguish the speaker's gestures and there can be significant individual and situational differences \cite{hostetter2011}. 

Additionally, coding schemes need to distinguish physical properties across semantically related sequences of gestures, i.e., ideational units\cite{calbris2011elements}. As discussed earlier, in an ideational unit, the physical properties of a gesture can be carried over to subsequent gestures. For example, Obama uses a beat gesture with a precision grip to emphasize a specific point and then holds that prevision grip as he does a sweep gesture to illustrate a related issue. The precision grip is not intrinsic to this sweeping gesture but rather is carried over from the previous beat gesture. During annotation, the annotator must ideally distinguish such carryovers of form. 

These shortcomings can be circumvented by not using annotations but instead prompting LLMs with just the context and utterance. However, the unconstrained gestural suggestions provided by the models can be hard to parse by machine making automated gesture generation on this basis difficult. Another approach is automatic annotation which would minimize laborious user input. However, this would also cede some degree of designer control. On the upside, we showed that leveraging an LLM's powerful generalization capability does not require a large amount of annotations. The important requirement that remains is to have sufficient annotations to span the speaker's gesture repertoire.

Another important issue is timing, specifically what parts of an utterance are accompanied by a co-speech gesture. Gestures are often associated with what the speaker seeks to emphasize \cite{clough_role_2020}. As a result, altering which part of an utterance is emphasized by a gesture can transform the meaning conveyed. It is therefore critical for co-speech gesture generation to derive emphasis information about the utterance. A common approach to this is to use prosodic cues such as pitch and pauses \cite{guellai_prosody_2014, fares2023}. However, this presumes that the spoken utterance that is driving gesture generation includes prosodic cues appropriate for the context in which the utterance occurs. A more flexible method is discourse analysis. For example, discourse analysis can identify the {\em rheme} of the utterance. This is the part of the utterance commenting on and providing new information about the topic or background of the conversation, what is referred to as the {\em theme}. The rheme tends to be more associated with gesturing \cite{mcneill1992hand,cassell2004beat}. It is likely that LLMs can also be leveraged to identify rhemes. 

However, in addition to temporal disconnects between gesture and speech, co-speech gestures often convey additional information that is not transmitted by the utterance at all. For example, in his speech, Obama says "I have sat in the Oval Office with both of the men who are running for president", accompanying the 'oval office' with a sweep of his right hand. This may simply function as an emphasis but, equally, alludes to the length of time that he has spent in that office -- 8 years -- and the experience he has therefore accumulated. It is a temporal sweep not conveyed by the semantic content of the speech and thus difficult to identify not only for humans but also for machine approaches.

Finally, the models used in this paper are proprietary and not open source. As a result, any update to the models may change the results without recourse for the researcher. This poses an issue for both reproducibility and reliability of any applications based on these models.

It is clear then from the work in this paper that LLMs hold great potential for gesture selection and, with that, generation. Yet, a number of challenges remain to be addressed. In the following we set out our current research efforts to formally and systematically explore the use of LLMs for gesture selection and implementation for gesture generation. 

\begin{enumerate}
\item Expansion of analysis to several types of speakers, including, for example, politicians, psychotherapists, and orators.
\item Annotation of these videos, going beyond the restricted set of gestures used in this paper and instead annotating all gestures.
\item Expansion of the types of prompts used to explore a) which prompts most effectively aid GPT in suggesting gestures and b) the minimal approach necessary to elicit appropriate and interpretable gestures.
\item Extension of analysis of the gestures produced to include rheme, theme, and ideational units. 
\item Evaluation of the appropriateness of the suggested gestures using human observer studies in addition to expert opinion.
\item Implementation of gesture generation by extracting gestures selected by GPT, matching to a gesture dictionary, and generating gestures for virtual human avatars.
\end{enumerate}


%% file: conclusion.tex
\section{Conclusion}
In this paper we showcased the use of LLMs to realize a generative mapping between verbal utterances and accompanying co-speech gestures. We explored approaches to using LLMs for gesture selection based on minimal to no example prompts. LLMs consistently suggested context-appropriate gestures. While this approach largely relies on laborious manual annotation, GPT required very few annotated examples to suggest suitable gestures. Nevertheless, it's important to note that the effectiveness of these gesture suggestions hinges on annotations derived from performances by a skilled speaker in a context and role that mirror the designer's intent for a virtual agent. Throughout our exploration of LLMs in this context, we not only highlighted their potent ability to select gestures but also underscored the challenges that still need to be surmounted for the complete implementation of LLMs for gesture generation. We outlined our ongoing research endeavors aimed at overcoming these obstacles.

In conclusion, we view this novel approach as a valuable and versatile tool for gesture researchers to use. However, we recognize that its suitability and alignment with goals may vary, as it either necessitates annotation (when example prompts are used) or interpretation (when example prompts are absent).  We nevertheless believe that it will be a fruitful avenue for many gesture researchers. 